\documentclass[prd,showpacs,preprintnumbers,nofootinbib]{revtex4}

\newcommand{\be}{\begin{equation}}
\newcommand{\beq}{\begin{equation}}
\newcommand{\ee}{\end{equation}}
\newcommand{\eeq}{\end{equation}}
\newcommand{\beqn}{\begin{eqnarray}}
\newcommand{\eeqn}{\end{eqnarray}}
\newcommand{\bea}[1]{\beq\begin{array}{#1}}
\newcommand{\eea}{\end{array}\eeq}
\newcommand{\eq}[1]{(\ref{#1})}

\newcommand{\sign}{\mathop{\rm sign}}

\newcommand{\dd}{{\mathrm d}}
\newcommand{\Z}{Z\!\!\! Z}

\newcommand{\diff}{\partial}

\newcommand{\NP}[3]{{Nucl. Phys. B}{\bf #1}, #3 (#2)}
\newcommand{\PL}[3]{{Phys. Lett. B}{\bf #1}, #3 (#2)}
\newcommand{\PRL}[3]{{Phys. Rev. Lett. }{\bf #1}, #3 (#2)}
\newcommand{\PR}[3]{{\em Phys. Rev. D}{\bf #1}, #3 (#2)}
\newcommand{\bbbone}{{{\mathchoice {\rm 1\mskip-4mu l} {\rm 1\mskip-4mu l}
{\rm 1\mskip-4.5mu l} {\rm 1\mskip-5mu l}}}}
\newcommand{\la}{\langle}
\newcommand{\ra}{\rangle}

\begin{document}

\preprint{MPI-PhT-2001-20}

\title{On the fermionic signature of the lattice monopoles}

\author{M.N. Chernodub}
\affiliation{Institute of Theoretical and  Experimental Physics,
B.Cheremushkinskaja 25, Moscow, 117259, Russia}

\author{V.I. Zakharov}
\affiliation{Max-Planck Institut f\"ur Physik, F\"ohringer Ring 6,
80805 M\"unchen, Germany}

\date{}

\begin{abstract}
We consider fermions in the field of
static monopole-like configurations in the Euclidean space-time.
In all the cases considered there exists an infinite number of zero modes,
labeled by frequency $i\omega$.
The existence of such modes is a manifestation of
instability of the vacuum in the presence of the monopoles and massless
fermions. In the Minkowski space the corresponding phenomenon is well
known and is a cornerstone of the theory of the magnetic catalysis.
Moreover, the well known zero mode of Jackiw and Rebbi corresponds to
the limiting case, $\omega = 0$. We provide arguments why the
chiral condensate could be linked to the density of the
monopoles in the infrared cluster. A mechanism which
can naturally explain the equivalence of the
critical temperatures for the deconfinement and chiral
transitions, is proposed.
We discuss possible implications for the phenomenology of the lattice
monopoles.
\end{abstract}


\maketitle

\section{Introduction }

Condensation of the monopoles is widely believed to be the confinement
mechanism.
It is natural then to try to reduce the chiral symmetry breaking to the
monopole physics as well. There are many numerical investigations of the
possible connection between the monopoles and chiral symmetry breaking, see, e.g.,
\cite{mon:lattice}.
On the theoretical side, the analysis proceeds usually along the lines
of the Banks-Casher criterion \cite{banks} which relates
the chiral symmetry breaking to the density of
the zero fermionic modes in a given bosonic background.
In case of the monopole-dominated vacuum, the elementary bosonic configuration
is
usually assumed to be a monopole-antimonopole pair.
The reason is that in the field of a monopole-antimonopole pair
there exist normalizable zero modes studied first in
Ref. \cite{stern}.

So far the properties of the fermions on the lattice were studied
mostly in the quenched approximation.
However, detailed measurements with dynamical fermions are imminent,
see, e.g., \cite{collaboration}.
In view of this, it is worth to revisit the problem of the fermions
in the monopole-dominated vacuum. In particular, we feel that it is
important to consider in more detail fermions in the field of
a single monopole, not of a monopole-antimonopole pair.
Indeed, if the QCD vacuum were dominated by magnetic dipoles
the standard explanation of the confinement would not work.
We consider three monopole-like configurations
which were introduced earlier.
We demonstrate that in all the cases there exists an infinite number of
solutions to the $d=4$ Dirac equation $D_{\mu}\gamma_{\mu}\psi(i\omega)=0$
in the Euclidean space-time labeled with imaginary frequency $i\omega$.
The solutions signal instability of the fermionic vacuum in the
presence of the magnetic monopoles. The instability is well known
in the language of the Minkowski space and is the starting point
of the theory of the monopole catalysis \cite{rubakov,callan,blaer}.

The instability of the fermionic vacuum in the presence of the monopoles
implies in fact inconsistency of the quenched approximation to study the
chiral symmetry breaking in the monopole-dominated vacuum. We will comment
on this in the conclusions.

\section{Fermionic modes}

\subsection{Equations}

In this section we rewrite the formalism of Jackiw and Rebbi
\cite{jackiw} in the Euclidean
space--time. We will study solutions of the Dirac equation:
\beqn
\gamma_\mu \Bigl(i \diff_\mu - \frac{1}{2} \tau^a A^a_\mu \Bigr) \Psi = 0\,,
\label{Dirac:equation}
\eeqn
where $\tau^a$ are the isospinor Pauli matrices and $\gamma_\mu$ are
the $4d$ Euclidean Dirac matrices:
\beqn
\gamma_0 = \left(\begin{array}{cc} 0 & -\bbbone \\ -\bbbone & 0 \\
\end{array} \right)\,,
\qquad
\vec \gamma = \Biggl( \begin{array}{cc} 0 & i \vec \sigma  \\
- i \vec \sigma & 0 \\ \end{array}
\Biggr)\,,
\nonumber
\eeqn
$\sigma_\alpha$ are the Pauli matrices and $\bbbone$ is the $2 \times 2$
unit matrix.

Moreover we will consider the static monopole gauge fields defined as,
\beqn
A^a_0 = n^a \, \Phi(r)\,, \quad A^a_i = \varepsilon^{aik}\, n^k \, A(r)\,,
\label{field:general}
\eeqn
where $n^i = x^i \slash r$ and $r^2 = x^2_i$. At large distances:
\beqn
 \quad \Phi(\infty) = \eta \ge 0\,, \quad
\lim\limits_{r\to \infty} r A(r) = -1 \,.
\label{boundary:conditions}
\eeqn
The case $\Phi\equiv 0$ and $A = - 1/r$ at all the distances corresponds to
the point-like Wu-Yang monopole ~\cite{wu}. Note that all the exact monopole solutions
of the pure Yang-Mills  equations are gauge copies of this monopole.
We will consider also fields regular at origin, $A(0)=0$.

We look for solutions of the Dirac equation \eq{Dirac:equation} of the following form:
\beqn
\label{fermi:ansatz}
\Psi_k(t,\vec x; \omega) = \frac{1}{\sqrt{2 \pi}} e^{- i \omega t} \,
\varphi_k (\vec x;\omega)\,,
\nonumber
\eeqn
where $\omega$ and $k$ are real,  continuous and discrete parameters,
respectively. The solutions obey the normalization condition,
\beqn\label{norm:4D}
\int \dd t \int \dd^3 x \, \Psi^\dagger_k (t, \vec x;\omega) \,
\Psi_{k'} (t, \vec x;\omega') = \delta(\omega - \omega') \, \delta_{k,k'}\,,
\eeqn
and the three--dimensional component of the zero mode is normalized as follows:
\beqn
\int \dd^3 x \, \varphi^\dagger_k (t, \vec x;\omega) \, \varphi_{k'} (t,
\vec x;\omega) =
\delta_{k,k'}\, .
\label{norm:3D}
\eeqn
The upper and lower components of the three
dimensional spinor field are denoted as:
\beqn
\varphi(\vec x) = \left(\begin{array}{c} \chi^+(\vec x)\\
\chi^-(\vec x) \end{array} \right)\,,
\label{psi:3D}
\eeqn
Following Ref.~\cite{jackiw} we regard the fields $\chi^\pm_{i\alpha}$ as
matrices. The spinor Pauli matrices act on this field as the matrix
multiplication, ${(\vec \sigma \chi^\pm)}_{i\alpha} = \vec \sigma_{ij}
\chi^\pm_{j\alpha}$, while the isospinor Pauli matrices act as follows:
${(\vec \tau \chi^\pm)}_{i\alpha} = \vec \tau_{\alpha \beta}
\chi^\pm_{j\beta} \equiv {(\chi^\pm \vec \tau^{tr})}_{i\alpha}$, where
$\vec \tau^{tr}$ is the transposed matrix, and $\vec
\tau^{tr} = \varepsilon \, {\vec \tau} \varepsilon $ with $\varepsilon = i
\tau^2$ being the totally antisymmetric tensor in two dimensions. In these notations,
the Dirac equation Eq.~\eq{Dirac:equation} becomes:
\beqn\label{Dirac:eq3}
\Bigl(- \omega \, M^\pm - \frac{1}{2} A^a_0 \, M^\pm \, \sigma^a\Bigr)
& \!\!\! \pm \!\!\! & \Bigl[(\vec \sigma, \vec \partial) M^\pm -
\frac{i}{2} A^a_i\, \sigma_i
\, M^\pm \, \sigma^a \Bigr] = 0\,,\\
M^\pm & \!\!\! = \!\!\! & \chi^\pm \, \varepsilon = g^\pm \cdot \bbbone +
g^\pm_a \cdot \sigma^a\,.
\label{M}
\eeqn
where $g^\pm$ and $g^\pm_a$ are unknown functions used to parameterize
the matrix $M^\pm$. We are looking for spherically
symmetric $s$--wave solutions of the equations,
$g^\pm = g^\pm(r)$
and $g^\pm_a = f^\pm(r) \, n_a$, where $f^\pm$ are scalar functions.
Substituting this
ansatz into Eq.~\eq{Dirac:eq3}, we
get two sets of differential equations:
\beqn
K^\pm f^\pm \mp \omega\, g^\pm = 0\,, \quad
D^\pm g^\pm \mp \omega\, f^\pm = 0\,,
\label{eqs:symbolic}
\eeqn
where
\beqn
K^\pm = \frac{\partial}{\partial r} + \frac{2}{r} + A(r) \mp \frac{1}{2}
\Phi(r)\,,
\quad
D^\pm = \frac{\partial}{\partial r} - A(r) \mp \frac{1}{2} \Phi(r)\,.
\label{KD}
\eeqn

We are mainly interested in the non--zero frequency case, $\omega \neq 0$,
so that the functions $f^\pm$ and $g^\pm$ are related to each other:
\beqn
f^\pm = \pm \frac{1}{\omega}\, D^\pm g^\pm\,.
\label{fg}
\eeqn
Four first order equations \eq{eqs:symbolic} are reduced then
to two differential equations
of the second order :
\beqn
K^\pm D^\pm g^\pm - \omega^2 \, g^\pm = 0\,.
\label{eq:2nd}
\eeqn
The solutions of these equations are discussed in the next subsection.

\subsection{Zero-mode solutions}

The normalizable fermionic modes can be readily found in case of the
point-like $\Z_2$ Wu--Yang monopole:
\beqn
\Psi_L(r;\omega) & = & \frac{{|\omega|}^{1/2}}{4\,\pi\,r}
\left(\begin{array}{c}
\Bigl[\bbbone - {\mathrm{sign}}(\omega) \, (\vec \sigma,\vec n)\Bigr] \,
\varepsilon \\
0\\
\end{array}\right) e^{ - i \omega t - |\omega| \, r} \,,\nonumber\\
~\label{psi:LR}\\
\Psi_R(r;\omega) & = & \frac{{|\omega|}^{1/2}}{4\,\pi\,r}
\left(\begin{array}{c}
0\\
\Bigl[\bbbone + {\mathrm{sign}}(\omega) \, (\vec \sigma,\vec n)\Bigr] \,
\varepsilon
\end{array}\right) \, e^{ - i \omega t - |\omega| \, r}\,,
\nonumber
\eeqn
where $\omega \in (-\infty,+\infty)$.


Next, let us consider a non-vanishing
$A_0$ component of the gauge field\footnote{In fact one should impose the condition
$\Phi(0)=0$. Otherwise, the field $A_0$ is not defined at the origin.
We assume that the transition from $\Phi=2\mu$ to $\Phi=0$ occurs at very small distances.},
$\Phi\equiv 2 \mu$. The zero modes can again be found explicitly and the functional form of
the solutions depends on
the value of $\omega$:
\beqn
\omega \le - \mu ~~~~~~~~~~&
\begin{array}{rcr}
\Psi_L(r;\omega) & = & \frac{\sqrt{-\mu - \omega}}{4\,\pi\,r}
\left(\begin{array}{c}
\Bigl[\bbbone + (\vec \sigma,\vec n)\Bigr] \,
\varepsilon \\
0\\
\end{array}\right) e^{ - i \omega t - (\omega + \mu) \, r} \,,\\
\Psi_R(r;\omega) & = & \frac{\sqrt{\mu - \omega}}{4\,\pi\,r}
\left(\begin{array}{c}
0\\
\Bigl[\bbbone - (\vec \sigma,\vec n)\Bigr] \,
\varepsilon
\end{array}\right) \, e^{ - i \omega t - (\omega - \mu) \, r}\,,
\end{array}
\label{mode:small}\\
~ & ~ \nonumber\\
- \mu < \omega < \mu ~~~~~~~~~~&
\begin{array}{rcr}
\Psi^{(1)}_R(r;\omega) & = & \frac{\sqrt{\mu + \omega}}{4\,\pi\,r}
\left(\begin{array}{c}
0\\
\Bigl[\bbbone + (\vec \sigma,\vec n)\Bigr] \,
\varepsilon \\
\end{array}\right) e^{ - i \omega t - (\omega + \mu) \, r} \,,\\
\Psi^{(2)}_R(r;\omega) & = & \frac{\sqrt{\mu - \omega}}{4\,\pi\,r}
\left(\begin{array}{c}
0\\
\Bigl[\bbbone - (\vec \sigma,\vec n)\Bigr] \,
\varepsilon
\end{array}\right) \, e^{ - i \omega t - (\mu - \omega) \, r}\,,
\end{array}
\label{mode:zero}\\
~ & ~ \nonumber\\
\omega > \mu ~~~~~~~~~~&
\begin{array}{rcr}
\Psi_L(r;\omega) & = & \frac{\sqrt{\omega - \mu}}{4\,\pi\,r}
\left(\begin{array}{c}
\Bigl[\bbbone - (\vec \sigma,\vec n)\Bigr] \,
\varepsilon \\
0\\
\end{array}\right) e^{ - i \omega t - (\omega - \mu) \, r} \,,\\
\Psi_R(r;\omega) & = & \frac{\sqrt{\omega + \mu}}{4\,\pi\,r}
\left(\begin{array}{c}
0\\
\Bigl[\bbbone + (\vec \sigma,\vec n)\Bigr] \,
\varepsilon
\end{array}\right) \, e^{ - i \omega t - (\omega + \mu) \, r}\,,
\end{array}
\label{mode:large}
\eeqn


The both cases considered so far correspond to  monopoles of zero size.
A famous example of a monopole with a core of finite size
is provided by the 't~Hooft--Polyakov~ monopole \cite{HP}
which involves also a Higgs field. In case of QCD the role of the Higgs field
is rather commonly ascribed to the $A_0$ component of the gauge field.
In particular,
such field configurations were
considered in Ref.~\cite{stern} in connection with the monopole physics and chiral symmetry breaking.

For analytical studies, it is convenient to consider the so called
Bogomol'ny limit ~\cite{Bogomolny} where the field configuration is
known explicitly:
\beqn
A(r) = \frac{2 \mu}{\sinh (2 \mu r)} - \frac{1}{r}\,, \quad
\Phi = \frac{2 \mu}{\tanh (2 \mu r)} - \frac{1}{r}\,,
\label{HP}
\eeqn
The gauge field is regular at the origin, $A(0) = \Phi(0) =0$.
The mass  parameter $2 \mu$ defines the monopole size and, simultaneously,
the value of the Higgs condensate at spatial infinity. This field
configuration is known as the BPS dyon~\cite{rossi}.

The solutions of Eqs.~(\ref{eqs:symbolic},\ref{KD},\ref{HP}) are
\beqn
g^\pm_i(r) & = & \frac{N}{r} \, {\Biggl[\frac{\sqrt{4 \mu^2 + \omega^2 \sinh^2 (2 \mu r)}
+ |\omega| \cosh 2 \mu r}{\sqrt{2}\,\mu} \Biggr]}^{(-1)^i |\omega| \slash (2 \mu)}
\nonumber\\
& & \cdot {\Biggl[
\frac{\cosh (2 \mu r) - \sqrt{4 \mu^2 + \omega^2 \sinh^2 (2 \mu r)}}{\cosh (2 \mu r)
+ \sqrt{4 \mu^2 + \omega^2 \sinh^2 (2 \mu r)}
}\Biggr]}^{(-1)^i \slash 2} \cdot
{\Biggl[\frac{\sinh( 2 \mu r)}{2 \mu r} \Biggr]}^{\pm \frac{1}{2}}\,,
\label{g:HP}
\eeqn
where the subscript $i=1,2$ corresponds to two independent solutions of
Eq.~\eq{eqs:symbolic}. The constant $N$ is defined by
the normalization condition \eq{norm:3D}.

The asymptotics of our solutions \eq{g:HP},
\beqn
\begin{array}{rcll}
g^{\pm}_i(r) & \propto & {(\mu r)}^{(-1)^i - 1} + O(\mu^2 r^2)\,,& r \to 0\,,\\
g^{\pm}_i(r) & \propto & \exp\Bigr\{ \frac{1}{2} (-1)^i \,
(\pm 1 + |\omega| \slash \mu )\,  r  \Bigl\}
+ O\Bigl(e^{- 2 \mu r}\Bigr)\,,& r \to \infty\,.
\end{array}
\eeqn
indicate that the $i=1$ solution is not normalizable at small
$r$ region while the
non--singular $i=2$ solution is normalizable. At large distances
the $g^+_2$ solution is always growing exponentially and thus not normalizable. However
the solution $g^-_2$ can be normalized provided
$|\omega| \le 1 \slash 2$.

Thus we get the following normalizable solution
\beqn
g^+ (r) = 0\,, \quad & &
g^- (r) = \frac{N}{r} \, {\Biggl[ \frac{\sqrt{4 \mu^2 + \omega^2 \sinh^2 (2 \mu r)}
+ |\omega| \cosh 2 \mu r}{\sqrt{2}\,\mu}
\Biggr]}^{\frac{|\omega|}{2\mu}} \nonumber \\
& & \cdot {\Biggl[ \frac{2\mu r}{\sinh (2\mu r)} \,
\frac{2 \mu \cosh (2\mu r) - \sqrt{4 \mu^2 + \omega^2 \sinh^2 (2 \mu r)}}{
2 \mu \cosh (2\mu r)
+ \sqrt{4 \mu^2 + \omega^2 \sinh^2 (2 \mu r)}}\Biggr]}^{\frac{1}{2}}\,.
\label{g:HP:norm}
\eeqn

The functions $f^\pm$ are defined by Eqs.~(\ref{fg},\ref{g:HP:norm}):
\beqn
f^+ (r) = 0 \,, \quad f^- (r) = \frac{1}{\omega \sinh ( 2 \mu r)} \Bigl(
\sqrt{4 \mu^2+ \omega^2 \sinh^2 (2 \mu r)} - 2 \mu \Bigr)\, g^- (r)\,.
\label{f:HP:norm}
\eeqn
Note that at large distances we recover the relation,
$f^- (r) = \sign (\omega) \, g^-(r)$, $cf.$ Eq.~\eq{psi:LR}.

Finally, combining Eqs.~(\ref{g:HP:norm},\ref{f:HP:norm})
we get the right--handed fermion zero mode:
\beqn
& & \Psi_R = \frac{N}{r} \, {\Biggl[ \frac{2\mu r}{\sinh (2\mu r)}
\frac{2 \mu \cosh (2\mu r) - \sqrt{4 \mu^2 + \omega^2 \sinh^2 (2 \mu r)}}{
2 \mu \cosh (2\mu r) + \sqrt{4 \mu^2
+ \omega^2 \sinh^2 (2 \mu r)}}\Biggr]}^{\frac{1}{2}} \label{HP:mode} \\
& & {\Biggl[ \frac{\sqrt{4 \mu^2 + \omega^2 \sinh^2 (2 \mu r)}
+ |\omega| \cosh 2 \mu r}{\sqrt{2}\,\mu} \Biggr]}^{\frac{|\omega|}{2\mu}}
\left(
\begin{array}{c}
0 \\
\Bigl[\bbbone + \frac{\sqrt{4 \mu^2 + \omega^2 \sinh^2 (2 \mu r)} - 2\mu}{
\omega \sinh (2 \mu r)}
\, (\vec \sigma,\vec n) \Bigr]
\varepsilon
\end{array}
\right)\,, \nonumber
\eeqn
where the frequency $\omega$ is restricted by the condition.
\beqn
\omega \le \mu\,.
\label{restriction}
\eeqn
This solution coincides (up to a gauge transformation) with the
fermion zero mode solution found in Ref.~\cite{simonov}.

The solutions ({\ref{mode:small}-\ref{mode:large},\ref{HP:mode})
can be linked to the fermion zero modes in the Georgi--Glashow model
coupled to the fermions\footnote{We do not describe this model and refer
reader to Ref.~\cite{jackiw} for details. The essential point is that
the Dirac equation for {\it static}
zero modes has the same form both in the $4d$ QCD and  $3d$
in the Georgi--Glashow model coupled to fermions (with identification
of the zero component of the gauge field in the former case with the
adjoint scalar field in the
latter).}. This model has been considered by Jackiw and Rebbi~\cite{jackiw}
who found the static fermion zero mode in the background of the
't~Hooft--Polyakov monopole:
\beqn
\Psi =
N\, \exp\Bigl\{\,
\int\limits^r_0 \dd r' \, \Bigl[A(r') - \frac{1}{2}
\Phi(r')\Bigr]\Bigr\} \cdot
\left(\begin{array}{c}
0\\
\varepsilon\\
\end{array}\right)
\,,\label{JR}
\eeqn
where the constant $N$ is determined from the three--dimensional
normalizability condition, Eqs.~(\ref{norm:3D},\ref{psi:3D}).

The link between our zero modes and the Jackiw--Rebbi solution \eq{JR}
can easily be established. First, let us consider the Wu--Yang monopole case,
Eqs.~({\ref{mode:small}-\ref{mode:large}). Setting $\omega=0$ we restrict ourselves
to  two solutions \eq{mode:zero}. The linear combination
$\Psi^{(1)}_R + \Psi^{(2)}_R$ coincides with the Jackiw--Rebbi mode
\eq{JR} while $\Psi^{(1)}_R - \Psi^{(2)}_R$ becomes its gauge copy.

The identification in the case of the 't~Hooft--Polyakov monopole,
Eq.~\eq{HP:mode},
is even more straightforward. Setting $\omega=0$ we get:
\beqn
\Psi =
\frac{1}{\mu} \sqrt{\frac{2\pi}{r\, \sinh(2 \mu r)}} \, \tanh(\mu r) \cdot
\left(\begin{array}{c}
0\\
\varepsilon\\
\end{array}\right)\,,
\nonumber
\eeqn
This expression coincides with the zero mode \eq{JR} where the fields
$A(r)$ and $\Phi(r)$ are given in Eq.~\eq{HP}.

\subsection{Perturbations on the potential}

The monopole field configurations which allow for exact zero-mode solutions
assume fixation of the gauge field $A_{\mu}$ at all the distances.
In reality of course one can hope to imitate the lattice monopole fields
only to some extent. In particular, there arise cuts off both
at large and small distances and the next question is, what
are the corresponding changes in the structure of the zero modes.
We address this question, on the qualitative level, in this subsection.
First, let us notice that
although in all the cases considered we found an infinite number
of the fermionic zero modes the status of these modes is somewhat different.
Namely in the first case, see Eqs.~(\ref{psi:LR}), there is symmetry between
the left- and right-handed modes. In the third case, on the hand, there
are right-handed modes alone, see Eq.~(\ref{HP:mode}). Finally, the modes
in
the second case considered, see Eqs.~(\ref{mode:small},\ref{mode:zero},
\ref{mode:large}), are of mixed nature.

The difference in the number of the right- and left-handed modes is
controlled in fact by the chiral anomaly:
\beqn
N_R-N_L~=~\int dt\int d^3r {{\bf H}^a\cdot {\bf E}^a\over 32\pi^2},
\label{anomaly}
\eeqn
where ${\bf H}^a and {\bf E}^a$ are color magnetic and electric fields,
respectively.
The crucial point is that the product $({\bf H}^a\cdot {\bf E^a})$
in the second and third examples considered in the previous
subsection does not disappear already on the classical level.
Note  that
to make use of (\ref{anomaly}) in our case one should introduce  a finite
range of integration over the time coordinate, $-T< t <T$ where T is large.

Now, if we modify the gauge field configurations, the difference
$N_R-N_L$ changes smoothly as far as the change in the r.h.s. of
Eq.~(\ref{anomaly}) is smooth. The corresponding analysis is trivial enough.

The situation is much more non-trivial in case of the Wu-Yang point-like
monopole. Namely let us introduce a cut off at small distances
so that $A_i \sim - 1/r$ only as far as $r>r_0$
while at short distances the potential vanishes, $A_i(0)=0$. Then
the zero modes found in the previous subsection
become non-renormalizable.  In other words, the zero modes
disappear altogether! To see this, it is convenient to use the
following relation:
\beqn
\int \dd^3 x \, \Biggr\{ {\Bigl(\tilde D g^\pm
\Bigr)}^2 + \Biggl[\omega^2 - \frac{1}{4} \Phi^2(r) \Biggr] \, g^{\pm,2}
\Biggr\} = 4 \pi \, \Bigl(r^2 g^{\pm,2}\Bigr) {\Biggl |}^\infty_0\,.
\label{rel:1}
\eeqn
where $\tilde D = \partial \slash \partial r - A(r)$
and independence of the functions $g^\pm$ on the angular variables is assumed.
Eq (\ref{rel:1} can be readily obtained by
multiplying the Eq.~(\ref{eq:2nd}) by the functions $g^\pm$,
integrating over the whole space and integrating by parts.
Eq.~(\ref{rel:1}) implies that $g\sim 1/r$ even if $A_i(0)=0$.
It follows then from Eq.~(\ref{fg}) that the function $f\sim 1/r^2$
at small $r$ and the zero-mode solution is not renormalizable.

We will comment on the physical meaning of this discontinuity in
the next section. Here we would like to notice only that the
introduction of the lattice spacing $a\neq 0$ allows to
introduce renormalizable solutions in any case. Moreover,
to be sensitive to the monopole field in the infrared we
should restrict ourselves to $\omega \ll 1/r_0$.
Then the normalization integral over the functions (\ref{psi:LR})
is dominated by $r\sim 1/\omega$. Upon the modification
of $A_i$ at small distances there emerges a new contribution
from the distances of order $a$. The new contribution does not
dominate provided that the product $(\omega \cdot r_0)(r_0/a)$
is small.

\section{Phenomenological applications. Conclusions}

The existence of an infinite number of the zero fermionic modes
indicates the instability of the fermionic vacuum in the presence
of the monopole-like field configurations. And, indeed, the instability
of the fermionic vacuum in the presence of the monopoles or dyons
is well known and is the starting point of the theory of
the monopole catalysis \cite{rubakov,callan,blaer}.
In particular, it is well known that the
S-wave interaction of massless fermions with Abelian monopoles is
anomalous in the sense that for some chiralities there exist only
coming-in waves while for other chiralities there exist only outgoing
waves, see discussion in \cite{rubakov1}.

In terms of this analogy, one can also easily understand the drastic effect
on the zero modes
of the modification of the Wu-Yang monopole field on the short distances,
see the subsection 2.3. Indeed, if one applies the Dirac equation to study
the motion of a massless fermion in the field of the 't Hooft-Polyakov monopole
then the result is that the fermion changes its charge due to the
W-boson exchange on the core of the monopole, see, e.g., \cite{rubakov1}.
As a result, the sign of the magnetic moment is changed as well
and the fermion but of opposite chirality
is emitted as a particle of the same chirality
with energy of order $1/r_0$. In our language, the modification
of the potential at arbitrary small distances leads to the concentration
of the wave function at these distances. As a result any weak interaction,
like interaction with $W$-bosons becomes crucial. Moreover, the energy of the
fermion is not conserved. Thus, the wave function becomes sensitive
only to the modifications of the monopole configuration at short distances
which are very difficult to describe realistically for the lattice monopoles.
Instead, we suggested (see subsection 2.3) to use the lattice
regularization and choice of $\omega$ to remain sensitive
to the monopole configuration at large distances.

The instability of the fermionic vacuum revealed through existence
of an infinite number of the zero modes implies that the results
of the numerical simulations in the quenched approximation and
with dynamical fermions differ substantially. While studying the
solutions of the Dirac  equation is sufficient to establish the
instability of the perturbative vacuum, it is much more difficult
to find the true fermionic vacuum in the presence of the monopoles.
To this end, one has to consider the full field theory.
This allows to include the effect of the anomaly
which arise at the quantum level. The simplification is that
because of the S-wave nature of the fermions the corresponding theory is
effectively two-dimensional \cite{rubakov,callan}.

The outcome of the calculations \cite{rubakov,callan} is that
fermionic condensate is formed around the monopoles:
\beqn
\la \bar \psi(r) \, \psi(r)\ra = \frac{1}{4 \pi r^2} f(r,t)\,,
\nonumber
\eeqn
where $r$ is the distance to the monopole center. The function $f$
was estimated in the leading order in Ref \cite{rubakov}:
\beqn
|f(r,t)| = \frac{1}{2 \pi r}\,.
\label{ff}
\eeqn
At least naively, one can think of this result as of a manifestation
of the Pauli principle: the decay of the vacuum is stopped once the
fermionic states which correspond to the fermions falling on the center
are occupied. Since the final field configuration \eq{ff} is static it
can be thought of as a Euclidean as well.

So far we have not discussed the question how much the monopole-like
configurations considered above resemble the lattice monopoles.
In particular, an obvious reservation is that we used
an approximation of a point-like Abelian monopole.
Such a bosonic field configuration would have an infinite action
and could not be important in lattice simulations (for review
and further references see, e.g., \cite{alive}). However, as the latest
measurements strongly indicate \cite{anatomy} the monopole size is much
smaller than the distance between the monopoles. As a result, the
point-like monopoles might in fact be a valid
approximation to interpret the results of the lattice simulations.

Reversing the question, we can say that by studying the properties
of the fermionic zero modes in the quenched approximation one can
independently judge how close the lattice monopoles to one or
another theoretical description. An advantage of this approach is
that it is gauge invariant.

The mechanism of the chiral symmetry breaking discussed in the
paper may also allow to predict the properties of the chiral
condensate in the confinement phase as well as to provide a link
between the chiral and deconfinement phase transitions. An
essential point in the Callan--Rubakov analysis of the appearance
of the chiral condensate near the static monopole lies in the fact
that the Abelian monopole is static. Obviously, the fermions need
a finite time to be attracted to the monopole core by the
interaction between the fermion spin and magnetic charge of the
monopole. As a result, short-lived monopoles ({\it i.e.} the
monopole--anti-monopole pairs collapsing in a short time) can not
be responsible for the chiral symmetry generation. The analogue of
the short living monopoles in the Euclidean space--time is a
small--sized monopole loops which are indeed observed in the
lattice simulations~\cite{cluster}. However, besides these short
monopole trajectories a large monopole cluster is present in the
confinement phase of the theory. The Minkowskian counterparts of
these lattice monopoles should serve as agents of the chiral
symmetry breaking according to our considerations above.

This picture naturally links the confinement and chiral phase
transitions. Namely, chiral condensate must be non--zero in the
presence of the large monopole cluster, {\it i.e.}, in the
confinement phase. In the high temperature, deconfinement phase,
the large monopole cluster is absent~\cite{cluster}, and, as a result,
the chiral symmetry breaking ceases to exist.

Naively, one may suggest that the chiral condensate is
proportional to the density of the largest monopole cluster,
$\rho_{IR}$, at
least, at zero temperatures,
\beqn
\langle \bar \psi \, \psi\rangle = C_\rho \, \rho_{IR}\,,
\label{psi:rho}
\eeqn
since, both quantities have the same dimension (mass${^3}$).
However, at not-zero temperatures the dependence should be more
general:
\beqn
\langle \bar \psi \, \psi\rangle = F(\rho_{IR},T)\,,
\eeqn
where the function $F$ obeys the property $F(0,T) = 0$. Suggestion
\eq{psi:rho} may be checked in future lattice simulations at
various lattice couplings.

In conclusion, let us summarize the finding of the present paper:

\begin{itemize}

\item In the quenched approximation one usually relies on the Banks-Casher
relation~\cite{banks} to study relevance of various bosonic field
configurations to the chiral symmetry breaking. However, we see that in
the quenched approximation the evaluation of the chiral condensate is
inconsistent in the presence of the monopole-like configurations because
of the instability of the fermionic vacuum. This inconsistency is another
manifestation of the phenomenon of the monopole catalysis
\cite{rubakov,callan,blaer}.

\item It would be very interesting to search numerically for solutions of
the Dirac equations (in the given gluon field background) corresponding to
the fermionic zero modes with imaginary frequencies, as discussed above.
If such solutions exist it would demonstrate independently the relevance
of the monopole-like configurations --- usually defined in an Abelian
gauge (see Ref.~\cite{reviews} for a review) --- in a non-Abelian gauge
theory. The advantage of such a search for the monopoles is that it does
not depend on the Abelian gauge explicitly.

\item Qualitatively, since the quenched approximation is in fact not
adequate to describe the fermionic vacuum the back reaction of the
fermions on the gluon fields can be unexpectedly strong. In particular,
extra monopole-antimonopole attraction is generated. Further analytical
studies of the effect are in progress now.

\item The chiral condensate seems to be linked with the density of the
monopoles in the largest monopole cluster. The proposed mechanism is very
attractive since it can naturally explain the equivalence of the
critical temperatures for the deconfinement and chiral transitions.

\end{itemize}

\begin{acknowledgments}
We have profited from discussion of the monopole physics with many
colleagues. Our special thanks are due to V.G.~Bornyakov,
F.V.~Gubarev, M.I.~Polikarpov, G.~Schierholz, A.Yu.~Simonov,
T.~Suzuki, P.~van~Baal, A.~Wipf. This work was partially supported
by grants RFBR 99-01230a, RFBR 01-02-17456, INTAS 00-00111 and
CRDF award RP1-2103. M.N.Ch.~acknowledges the kind hospitality of
the staff of the Max-Planck Institut f\"ur Physik (M\"unchen).
\end{acknowledgments}

\end{document}